
\magnification=1200
\hsize 15.0 true cm
\vsize 23.0 true cm
\def\c{\centerline}
\def\v{\vskip 1pc}
\def\ej{\vfill\eject}
\def\r{\vec r}
\def\ra{\rangle}
\def\la{\langle}

\def\tr{{\rm tr}}
\def\half{{1 \over 2}}
\def\da{\dagger}
\def\pup{{\rm p} \uparrow}
\overfullrule 0 pc
\
\vskip 1.5pc
\parindent 3pc\parskip 1pc
\c{\bf Proton spin content from skyrmions}
\vskip 4 pc
\c{G. K\"albermann$^{\rm a}$, J.M. Eisenberg$^{\rm b,c}$,
Andreas Sch\"afer$^{\rm d}$}
\v
\c{$^{\rm a}$ {\it Rothberg School for Overseas Students and Racah Institute
of Physics}}
\c{\it Hebrew University, 91904 Jerusalem, Israel}
\vskip 0.5 pc
\c{$^{\rm b}$ {\it School of Physics and Astronomy}}
\c{\it Raymond and Beverly Sackler Faculty of Exact Sciences}
\c{\it Tel Aviv University, 69978 Tel Aviv, Israel}
\vskip 0.5 pc
\c{$^{\rm c}$ {\it TRIUMF, 4004 Wesbrook Mall, Vancouver, B.C., Canada V6T
2A3}}
\vskip 0.5 pc
\c{$^{\rm d}$ {\it Institut f\"ur Theoretische Physik der Universit\"at
Frankfurt am Main}}
\c{\it Postfach 11 19 32, D-60054 Frankfurt, Germany}

\vskip 4pc
\noindent {\bf Abstract:-}  It is well known that in lowest order the
skyrmion model of the nucleon gives vanishing spin content.  With new data
indicating a proton spin content $\Delta\Sigma = 0.22\pm 0.14,$ it
is an increasing challenge to find ways in which the skyrmion can move
away from the null
result.  We show here that a particular term in the skyrmion lagrangian
in SU(3) involving six derivatives of the field can, with plausible
parameters, yield a spin content consistent with present experiment.

\vfill

\noindent August, 1994.

\ej

\baselineskip 15 pt
\parskip 1 pc \parindent 3pc

Of the various models applied to the explanation of the original EMC
measurement [1] of the spin content of the proton, it remains the
distinction of the skyrmion that it implies [2] in its lowest-order
statement a small---indeed, a vanishing---value for that quantity.
(Reviews of the current general status of the spin-content problem are
given in Refs. [3,4].)  This, however, presented the Skyrme model [5]
with a special challenge when the SMC experiment [6] recently found a
spin content of $\Delta\Sigma = 0.22\pm 0.14.$  It has not proved easy,
despite several attempts [7,8]---and notably the sustained effort of the
group centered about Syracuse [9-11]---to move the Skyrme model away
from its statement of vanishing, or very small, spin content.
On the other hand the spin structure of the nucleon should allow a
crucial test of the Skyrme model, which fits a large number of hadronic
data successfully, but an important feature of which, namely, the
specific linking of spin and isospin and its consequences for spin
content, has not yet been fully assessed.  If $\Delta\Sigma\approx 0$
is really an indisputable property of the Skyrme model in its general
form, the spin data could rule it out definitively. The present paper
shows that this is not the case and identifies a source of
nonnegligible spin.  This is found in the form of a term in an extended
Skyrme lagrangian involving six derivatives of the field.  In SU(3) the
particular form we use is different from that considered in the past
[12-14] for the stabilization
of the skyrmion through an $\omega$-like exchange, and, with
reasonable parameters, leads to a spin content near that of experiment.
This result may be viewed, in some sense, as a
simpler, effective version of the Syracuse studies [9-11], which are
based on a large number of couplings to various mesons.

The skyrmion is here described by the lagrangian
$$\eqalign{{\cal L} & = {\cal L}_2 + {\cal L}_{4} + {\cal L}_{4,1}
+ {\cal L}_{4,2} + {\cal L}_{6,1} + {\cal L}_{6,2} + {\cal L}_{SB}\cr
& = -{F_\pi^2 \over 16} \tr(L_\mu  L^\mu )
+ {1 \over 32 e^2} \tr[L_\mu,L_\nu]^2 \cr
& + {\gamma_1 \over 4 e^2} \tr (L_\mu L^\mu L_\nu L^\nu)
+ {\gamma_2 \over 8 e^2} \tr (L_\mu L^\mu) \tr (L_\nu L^\nu) \cr
& - \epsilon_1 {g_\omega^2\over m_\omega^2} \tr (B^\mu B_\mu)
- \epsilon_2 {g_\omega^2\over 2 m_\omega^2} \tr (B^\mu) \tr ( B_\mu) \cr
& + \bigg[{F_\pi^2 \over 32}(m_\pi^2 + m_\eta^2) \tr (U + U^\da - 2)
+ {\sqrt{3}F_\pi^2 \over 24}(m_\pi^2 - m_K^2) \tr (\lambda_8(U + U^\da))
\bigg],}
\eqno(1)$$
apart from an anomalous contribution to the $\eta'$ mass which is not
relevant for our purposes here.
In the following we shall systematically refer to the terms
by their corresponding
subscripted forms as given in the first line of eq. (1).  Here
$L_\mu \equiv U^\da\partial_\mu U$ and
$$B^\mu \equiv -{\epsilon^{\mu\alpha\beta\gamma} \over 24\pi^2}
L_\alpha L_\beta L_\gamma,\eqno(2)$$
where $U(\r,t)$ is the U(3) chiral field, $F_\pi$ is the
pion decay constant (with experimental value 186 MeV), and $e$ is
the Skyrme parameter.

Let us motivate our choice for the lagrangian in some detail:  To
describe the nucleon spin structure correctly even only at the
qualitative level one should include terms which correspond to
$\omega$-exchange because the three-vector part of the $\omega$ couples
to spin. As $B_{\mu}$ relates to the baryon current one consequently
has to include terms quadratic in it. We shall show that the term
${\cal L}_{6,1},$ which contains only a single trace over field
variables, modifies the spin content of the skyrmion substantially.  In
contrast ${\cal L}_{6,2},$ which is a product of two such traces, does
not affect the spin (in line with the suggestion [9-11]
of the OZI rule).  Still higher-order terms or other sixth-order
terms might modify the results quantitatively but not qualitatively; we
therefore disregard them for the time being.  We shall show that
${\cal L}_{6,1}$ generates naturally a nonzero skyrmion spin and this
suggests that any complete treatment of the skyrmion will get
$\Delta\Sigma \neq 0.$

Since we take sixth-order terms into account we also add the two
fourth-order terms ${\cal L}_{4,1}$ and ${\cal L}_{4,2}$ for
completeness.  The first of these corresponds to a term which Ryzak [7]
claimed to give a $\Delta\Sigma$ substantially different from zero, a
claim our results do not support.  These terms are notorious
as they contain quartic time derivatives and also work to destabilize the
skyrmion. No consensus exists on their interpretation and treatment. We
regard it therefore as supportive of the straightforward interpretation
of the Skyrme model that these terms are unimportant for $\Delta\Sigma$
in our calculation.

The parameters $\gamma_1$ and $\gamma_2$ in eq. (1)
can be related to $\pi\pi$ scattering, and the terms ${\cal L}_{4,1}$ and
${\cal L}_{4,2}$ have been used in the past to
increase $NN$ attraction in the skyrmion (see [15] and, e.g., the review
in Ref. [14]).  The parameters
$\epsilon_1 g_\omega^2/m_\omega^2$ and $\epsilon_2 g_\omega^2/m_\omega^2$
are coefficients of six-derivative terms
${\cal L}_{6,1}$ and ${\cal L}_{6,2},$ respectively, both of which
have been used in the past as possible $\omega$-coupling repulsive terms
for stabilizing the skyrmion [12,13].  They are equivalent [13] in
SU(2), but not in SU(3).  (Terms with four derivatives that are
equivalent in SU(2) but different in SU(3) are, of course, well known
[16].)  (In the past, a contribution to spin content from a
term that would
correspond in this classification to an eight-derivative part has been
considered [8] and found to give a small contribution,
but, to the best of our knowledge, the six-derivative term has
thus far been bypassed.)  The forms of the
coefficients of ${\cal L}_{6,1}$ and ${\cal L}_{6,2}$ are selected to
allow easy comparison with the $\omega NN$ coupling constant.
We keep $\epsilon_1 + \epsilon_2 = 1,$ absorbing overall strength
into $g_\omega^2,$ with $m_\omega,$ the mass of the $\omega,$
fixed at its experimental value, so that
${\cal L}_{6,1} + {\cal L}_{6,2}$ at the SU(2) level continues to give
the usual $\omega$-like coupling.  The flavor symmetry-breaking term
${\cal L}_{SB}$ in eq. (1) is well known [17] to be important for work
with the SU(3) skyrmion.

In order to generate the proton spin content from ${\cal L}$ of eq. (1),
we introduce the U(3) matrix
$$U = \exp\bigg[{2 i \over F_\pi}\bigg(\eta'+ \sum_{a=1,8} \lambda_a
\phi_a\bigg)\bigg],\eqno(3)$$
where $\phi_a$ is the pseudoscalar octet and $\eta'$ is the ninth
pseudoscalar meson.  As is well known [7,8], there will be no
contribution to the spin content from a U(1) axial current that is a
complete four-derivative, since this vanishes in producing $\Delta\Sigma$
as an integral over all space.  We construct the axial current of
interest to us here out of a term in the lagrangian of the form [8]
$${\cal L}' = (2/F_\pi) \partial_\mu\eta' J^\mu\eqno(4)$$
generated by varying $\eta'.$  In the evaluation of this $J^\mu,$ we use
the customary approach of collective coordinates for the time dependence [18]
$$U(\vec r,t) = A(t) U_0(\vec r) A^\da(t),\eqno(5)$$
with the hedgehog embedded in SU(3) as
$$U_0 = \exp[i \vec\lambda\cdot\vec r F(r)],\eqno(6)$$
where $F(r)$ is the profile function.  It is well known [8] that the
static hedgehog contribution to spin content vanishes on grounds of
grand-spin symmetry, so that a nonzero value is obtained only through the
collective-coordinate rotation.
In evaluating the nonstatic component $L_0$ with quantization of the
collective-coordinate transformation we encounter
$$A \dot A^\da = {i \over 2 \alpha^2}\vec\lambda\cdot\vec R
+ {i \over 2 \beta^2} \sum_{a=4,7} \lambda_a R_a
+ \dot q_i C_{i8}(q)\lambda_8,\eqno(7)$$
where $q_i$ is an SU(3) variable and the last term does not contribute in
our expressions.  The quantities $\alpha^2$ and $\beta^2$ relate in the
usual way to the SU(3) moments of inertia [17], and $R_a,\ a =
1,2,\dots,8,$ are the ``right'' SU(3) generators.

Since the terms ${\cal L}_{4,2}$ and ${\cal L}_{6,2}$ have been treated
extensively in the literature, and, more to the point, do not contribute
to the flavor-singlet axial current, we shall not repeat the well known
results for them here.  For ${\cal L}_{4,1}$ and ${\cal L}_{6,1}$ the
new, additional static contributions to the skyrmion mass are
$$M_{4,1} = -4\pi{\gamma_1 \over 2 e^2} \int_0^\infty r^2 dr
\bigg(F'^2 + 2{\sin^2 F \over r^2}\bigg)^2,\eqno(8)$$
and
$$M_{6,1} = 288\pi \bigg({1 \over 24 \pi^2}\bigg)^2
 {\epsilon_1 g_\omega^2 \over m_\omega^2}
\int_0^\infty r^2 dr
\bigg(F' {\sin^2 F \over r^2}\bigg)^2.\eqno(9)$$
The corresponding new contributions to the total moment-of-inertia
parameters $\alpha^2$ and $\beta^2$ are
$$\alpha_{4,1}^2 = -16 \pi {\gamma_1 \over 3 e^2}
 \int_0^\infty r^2 dr \sin^2 F
\bigg(F'^2 + 2{\sin^2 F \over r^2}\bigg),\eqno(10)$$
$$\beta_{4,1}^2 = -4 \pi{\gamma_1 \over e^2} \int_0^\infty r^2 dr
\sin^2 {F \over 2} \bigg(F'^2 + 2{\sin^2 F \over r^2}\bigg),\eqno(11)$$
and
$$\alpha_{6,1}^2 = 384 \pi
\bigg({1 \over 24 \pi^2}\bigg)^2
 {\epsilon_1 g_\omega^2 \over m_\omega^2}
\int_0^\infty r^2 dr
\bigg({F'\sin^2 F \over r}\bigg)^2,\eqno(12)$$
$$\beta_{6,1}^2 = 64 \pi \bigg({1 \over 24 \pi^2}\bigg)^2
 {\epsilon_1 g_\omega^2 \over m_\omega^2}
\int_0^\infty r^2 dr \sin^2 {F \over 2} {\sin^2 F \over r^2}
\bigg(2 F'^2 + {\sin^2 F \over r^2}\bigg),\eqno(13)$$

The term ${\cal L}_{4,1}$ yields a contribution to the matrix
element of the flavor-singlet axial current given by
$$\la\pup|J_{4,1}^3|\pup\ra = -2 \pi {\gamma_1 \over 3 e^2}
\int_0^\infty r^2 dr
\bigg[{1 \over 4 \alpha^4} F' \sin^2 F
+ {1 \over \beta^4} \bigg(F' - 2{\sin F \over r}\bigg)
\sin^2 {F \over 2}\bigg] ;\eqno(14)$$
this expression differs from that in [7] and leads to a
smaller contribution to spin content for ${\cal L}_{4,1}.$  The
corresponding matrix element for ${\cal L}_{6,1}$ is
$$\eqalign{\la\pup|J_{6,1}^3|\pup\ra & = {32 \pi \over 3}
\bigg({1 \over 24 \pi^2}\bigg)^2 {\epsilon_1 g_\omega^2 \over m_\omega^2}
 {1 \over \beta^4} \cr
& \times \int_0^\infty r^2 dr \sin^2 {F \over 2} \sin F
 {1 \over r} \bigg(F'^2 - F' {\sin F \over r}
+ {\sin^2 F \over r^2}\bigg).}\eqno(15)$$
As has been pointed out [7], these calculations are ``somewhat tedious
but otherwise straightforward,'' and so to assure their reliability we
have evaluated eqs. (10) through (15) both by hand and using a
symbol-manipulation program.  In arriving at these results, we have also
used the feature of the ``right'' SU(3) algebra that $R_4^2 + R_5^2 -
R_6^2 - R_7^2 = R_3,$ and $\la\pup|R_3|\pup\ra = -\half,$ as well as
the identification of the spin content through
$\Delta\Sigma = 2 \la\pup|J^3|\pup\ra.$
In $J_{6,1}^3$ only symmetric combinations of the $R_a$s appear.
On the other
hand, $J_{4,1}^3$ contains both commutators and anticommutators.  We
drop the former on the basis of arguments [9] that in going from
the classical to the quantal results for the rotation operators one
must symmetrize them, whence the commutators do not contribute.
This eliminates the term in the integrand of eq. (14) with the
coefficient $1/\alpha^4$ and also divides the right-hand side of that
equation by 2.  Since the contribution of $J_{4,1}^3$ is in any event
not very appreciable,
our conclusions below are not changed significantly by this procedure.

While the choice of some of the parameters appearing in the
lagrangian of eq. (1) is straightforward, others are
not so easily fixed.  The situation is eased somewhat by the realization
that the contribution of the ${\cal L}_{4,1}$ term to the spin content is
very small, on the order of 0.04 for a maximizing case with $\gamma_1 \sim
0.15$ and $e = 5$; efforts further to enhance that contribution led to a
breakdown in the stability of the skyrmion solution.  Thus, in the
following, we drop ${\cal L}_{4,1}$ as well as ${\cal L}_{4,2},$ which
also eliminates concern over terms in the lagrangian with higher than
bilinear time derivatives.  We then fix
$g_\omega^2/4\pi = 10,$ as implied by [13] $\omega NN$ coupling and by [12]
the decay $\omega \rightarrow 3\pi$.  Unfortunately,
there seems to be no very direct way to determine the crucial coefficient of
the ${\cal L}_{6,1}$ term, $\epsilon_1.$  The natural candidate would be to
compare with $\eta' \rightarrow 5\pi,$ but this vanishes dynamically
 for ${\cal L}_{6,1},$ and indeed experimentally [19] the branching ratio
for this mode seems to be less than a percent or so.  A transition that does
not vanish dynamically is $\eta' \rightarrow \eta + 4 \pi,$ but this,
of course, is kinematically forbidden as a decay.
Thus we proceed by fixing $F_\pi$ and $e$ so as to produce reasonable values
for nucleon properties and seeking a range for $\epsilon_1$ that yields
sizable spin content.  With $F_\pi = 130$ MeV and $e = 20$, we have
$M_N = 952$ MeV and $M_\Delta = 1,241$ MeV.
The SU(2) result for the axial coupling constant is $g_A = 1.21,$
in surprisingly good
agreement with experiment given the usual experience with skyrmions [18].
The nucleon charge radius is 0.71 fm, and the magnetic moments are
$\mu_{\rm p} = 2.04$ and $\mu_{\rm n} = -1.24.$
We then find that $\epsilon_1 = -0.7$ yields a spin content of
$\Delta\Sigma = 0.24.$ The centroid
of the masses of particles with nonzero strangeness then reaches
roughly 1,650 MeV, after subtraction of the zero-point energy [17].
Compensation is possible by reducing $F_\pi,$ but
further adjustment of parameters does not seem worthwhile in the face of
the poorly known value for $\epsilon_1.$
As a rough indicator of sensitivities, varying $\epsilon_1$ between -0.6
and -0.8 causes $\Delta\Sigma$ to range between 0.18 and 0.32 and the
strange-mass centroid to change between 1,610 MeV and 1,670 MeV.
The reason for the rather large contribution to the spin content from
the ${\cal L}_{6,1}$ term lies in the reduction of the total $\beta^2$
caused by the negative $\beta_{6,1}^2.$  Thus the ${\cal L}_{6,1}$
term shows a possible source of spin content for the skyrmion well within
the range of present experiment, but further precision in the theory
requires a better determination of its coefficient.
\v\v
This research was supported in part by the Israel Science Foundation
and in part by the Yuval Ne'eman Chair in Theoretical Nuclear Physics
at Tel Aviv University.  We are grateful to Avraham Gal, Byron Jennings,
Ami Leviatan, and Jechiel Lichtenstadt for discussions pertaining to it.

\vskip 1 cm

\noindent {\bf References:-}
\v
\baselineskip 12pt
\parskip 0pc
\parindent 1pc
\hangindent 2pc
\hangafter 10

\item{1.}  EMC, J. Ashman et al., Phys. Lett. B 206 (1988) 364; Nucl. Phys.
B 328 (1989) 1.
\v
\item{2.}  S.J. Brodsky, J. Ellis, and M. Karliner, Phys. Lett. B 206
(1988) 309.
\v
\item{3.}  J. Ellis and M. Karliner, Phys. Lett. B 313 (1993) 131 and
``Spin structure functions,'' to be published in Proc. Thirteenth
Int. Conf. on Particles and Nuclei, PANIC '93, Perugia, Italy, June, 1993.
\v
\item{4.}  F.E. Close and R.G. Roberts, Phys. Lett. B 316 (1993) 165 and
F.E. Close, ``The nucleon spin crisis bible,'' talk given at the Workshop
on Qualitative Aspects and Applications of Nonlinear Evolution Equations,
Trieste, Italy, May, 1993.
\v
\item{5.}  T.H.R. Skyrme, Proc. Roy. Soc. London, Series A, 260 (1961)
127; 262 (1961) 237 and Nucl. Phys. 31 (1962) 556.
\v
\item{6.} SMC,  D. Adams et al., Phys. Lett. B 329 (1994) 399.
\v
\item{7.}  Z. Ryzak, Phys. Lett. B 217 (1989) 325; B 224 (1989) 450.
\v
\item{8.}  T.D. Cohen and M.K. Banerjee, Phys. Lett. B 230 (1989) 129.
\v
\item{9.}  R. Johnson, N.W. Park, J. Schechter, V. Soni, and H. Weigel,
Phys. Rev. D 42 (1990) 2998.
\v
\item{10.}  J. Schechter, A. Subbaraman, and H. Weigel,
Phys. Rev. D 48 (1993) 339.
\v
\item{11.}  J. Schechter, V. Soni, A. Subbaraman, and H. Weigel,
Mod. Phys. Lett. A 7 (1992) 1.
\v
\item{12.}  G.S. Adkins and C.R. Nappi, Phys. Lett. B 137 (1984) 251.
\v
\item{13.}  A. Jackson, A.D. Jackson, A.S. Goldhaber, G.E. Brown, and
L.C. Castillejo, Phys. Lett. 154B (1985) 101.
\v
\item{14.}  J.M. Eisenberg and G. K\"alberman, Prog. Part. Nucl. Phys. 22
(1989) 1.
\v
\item{15.}  J.F. Donoghue, E. Golowich, and B.R. Holstein, Phys. Rev.
Lett. 53 (1984) 747.
\v
\item{16.} G. Pari, B. Schwesinger, and H. Walliser, Phys. Lett. B 255
(1991) 1.
\v
\item{17.}  H. Yabu and K. Ando, Nucl. Phys. B 301 (1988) 601.
\v
\item{18.} G.S. Adkins, C.R. Nappi, and E. Witten, Nucl. Phys. B 228
(1983) 552.
\v
\item{19.}  Particle Data Group, K. Hikasa et al., Phys. Rev. D 45 (1992)
S1.
\bye